\documentclass[conference,a4paper]{IEEEtran}
\usepackage{amsmath,amssymb,amsfonts}
\usepackage{graphicx}
\usepackage{booktabs}
\usepackage{microtype}

\usepackage{caption}
\usepackage{subcaption}
\usepackage{pbalance} % Automatically balances last page

\usepackage{pgfplots}
\pgfplotsset{compat=1.18}

\usepackage{enumitem}
\usepackage{acronym}

\usepackage{xcolor}

\usepackage[hidelinks]{hyperref}
\usepackage[nameinlink]{cleveref}
\usepackage{orcidlink}

\title{Estimating SSIM from  MSE\\ for DCT-Based Compressed Images via Modeling Local Error Statistics}
\author{
Luc Trudeau\,\orcidlink{0000-0002-4292-5875}$^{1}$
\qquad
Maria G. Martini\,\orcidlink{0000-0002-8710-7550}$^{2}$\\
$^{1}$ Université du Québec à Rimouski, Rimouski, QC, Canada\\
$^{2}$ Kingston University London, London, UK
}

\begin{document}

\acrodef{A/V}	{Audio/Video}
\acrodef{AOM}	{Alliance for Open Media}
\acrodef{AVC}	{Advanced Video Coding}
\acrodef{AWS}	{Amazon Web Services}
\acrodef{bpc}	{bits per channel}
\acrodef{BQoE}	{Behavior Quality of Experience}
\acrodef{CBR}   {Constant Bitrate}
\acrodef{CABAC}	{Context-Adaptive Binary Arithmetic Coding}
\acrodef{CAVLC}	{Context-Adaptive Variable-Length Coding}
\acrodef{CDN}	{Content Delivery Network}
\acrodef{CDR}	{Continuous Dynamic Range}
\acrodef{CfE}	{Call for Evidence}
\acrodef{CIE}	{International Commission on Illumination}
\acrodef{CQoE}	{Cognitive Quality of Experience}
\acrodef{CRF}	{Constant Rate Factor}
\acrodef{CTU}	{Coding Transform Unit}
\acrodef{DASH}	{Dynamic Adaptive Streaming over HTTP}
\acrodef{DCT}  {Discrete Cosine Transform}
\acrodef{DLM}	{Detail Loss Metric}
\acrodef{ECDF}  {Empirical Cumulative Distribution Function}
\acrodef{ECG}	{Electrocardiogram}
\acrodef{EEG}	{Electroencephalogram}
\acrodef{EPG}	{Electronic Program Guide}
\acrodef{FHD}	{Full HD}
\acrodef{GOP}	{Group of Pictures}
\acrodef{GPU}	{Graphics Processing Unit}
\acrodef{HAS}	{HTTP Adaptive Streaming}
\acrodef{HD}	{High Definition}
\acrodef{HDR}	{High Dynamic Rendering}
\acrodef{HEVC}	{High Efficiency Video Coding}
\acrodef{HLS}	{HTTP Live Streaming}
\acrodef{HM}	{HEVC Test Model}
\acrodef{HLS}	{HTTP Live Streaming}
\acrodef{HVS}	{Human Visual System}
\acrodef{HW}	{Hammerstein-Wiener}
\acrodef{IP}	{Internet Protocol}
\acrodef{IPR}	{Intellectual Property Rights}
\acrodef{IQA}	{Image Quality Assessment}
\acrodef{JPEG} {Joint Photographic Experts Group}
\acrodef{JND}	{Just Noticeable Difference}
\acrodef{LPIPS} {Learned Perceptual Image Patch Similarity}
\acrodef{LTE}	{Long Term Evolution}
\acrodef{MOS}	{Mean Opinion Score}
\acrodef{MPEG}	{Moving Pictures Expert Group}
\acrodef{MSE}	{Mean Square Error}
\acrodef{MSSIM} {Mean Structural Similarity Index}
\acrodef{MV}	{Motion Value}
\acrodef{NARX}	{Nonlinear Autoregressive Network with Exogenous Inputs}
\acrodef{OTT}	{Over The Top}
\acrodef{OS}	{Operating System}
\acrodef{PLCC}	{Pearson Linear Correlation Coefficient}
\acrodef{PLR}	{Packet Loss Rate}
\acrodef{POP}	{Point of Presence}
\acrodef{PSNR} {Peak Signal to Noise Ratio}
\acrodef{QoE}	{Quality of Experience}
\acrodef{QoS}	{Quality of Service}
\acrodef{QP}	{Quantization Parameter}
\acrodef{RAN}	{Radio Access Network}
\acrodef{RD}	{Rate-Distortion}
\acrodef{RMSE}	{Root Mean Square Error}
\acrodef{ROI}	{Region of Interest}
\acrodef{RR}	{Reduced Reference}
\acrodef{RTMP}	{Real Time Messaging Protocol}
\acrodef{RTT}	{Round Trip Time}
\acrodef{SAO}	{Sample Adaptive Offset}
\acrodef{SC}	{Spatial Complexity}
\acrodef{SD}	{Standard Definition}
\acrodef{SDR}	{Standard Dynamic Range}
\acrodef{SI}	{Spatial Information}
\acrodef{SoP}	{Sense of Presence}
\acrodef{SQI}	{Streaming Quality Index}
\acrodef{SROCC}	{Spearman's Rank Correlation Coefficient}
\acrodef{SSIM}	{Structural Similarity Index}
\acrodef{STSQ}	{Short Term Subjective Quality}
\acrodef{SVC}	{Scalable Video Coding}
\acrodef{SVM}	{Support Vector Machine}
\acrodef{SVR}	{Support Vector Regression}
\acrodef{TCP}	{Transmission Control Protocol}
\acrodef{TC}	{Temporal Complexity}
\acrodef{TI}	{Temporal Information}
\acrodef{TMO}	{Tone Mapping Operator}
\acrodef{TVSQ}	{Time Varying Subjective Quality}
\acrodef{UHD}	{Ultra High Definition}
\acrodef{UDP }	{User Datagram Protocol}
\acrodef{VCI} 	{Video Complexity Index}
\acrodef{VMAF}  {Video Multimethod Assessment Fusion}
\acrodef{VoD}	{Video on demand}
\acrodef{VQA}	{Video Quality Assessment}
\acrodef{VQMT}	{Video Quality Measurement Tool}
\acrodef{VBR}	{Variable Bit Rate}
\acrodef{VIF}	{Visual Information Fidelity}
\acrodef{VIFP}	{Visual Information Fidelity - Pixel Domain}
\acrodef{VR} 	{Virtual Reality}
\acrodef{WCG}	{Wide Color Gamut}

\maketitle
\begin{abstract}
Efficient and perceptually meaningful quality assessment is a fundamental requirement for image and video processing, compression, and streaming systems. This article shows that, in the context of \ac{DCT}-based compressed images, \ac{SSIM} can be approximated from global \ac{PSNR} or \ac{MSE} using local statistics derived only from the reference image. While prior work assumes access to local \ac{MSE}, we propose two approaches to approximate local \ac{MSE} by redistributing the global \ac{MSE} using variance or standard-deviation-based weighting. Experiments on the Kodak and Xiph Subset1 datasets across a range of JPEG quality levels demonstrate that both approaches provide accurate and robust \ac{SSIM} approximations, substantially outperforming the global \ac{MSE} baseline. The proposed framework is designed to extend naturally to video, where reference-derived statistics can be amortized across multiple encodes of the same content.
\end{abstract}
\section{Introduction}
\label{sec:intro}

Video and image compression requires objective quality metrics that are both efficient and perceptually meaningful. While subjective quality assessment remains the ground truth, it is often prohibitively expensive or impractical to use in many scenarios. Full reference objective quality metrics are mathematical models and algorithms designed to compare an image or block $x$ and its distorted counterpart $y$.

\ac{PSNR}, derived directly from \ac{MSE}, is the most widely used objective quality metric in codec development due to its simplicity and low computational cost. \ac{PSNR} expresses reconstruction fidelity on a logarithmic decibel scale, emphasizing relative error differences while remaining agnostic to image content and spatial structure.
However, \ac{PSNR} correlates poorly with perceived quality, motivating established and widely used alternatives such as SSIM~\cite{wang2004image}, which remains simple and interpretable despite not representing the state of the art in image quality assessment:
\begin{equation}
\label{eq:ssim}
\text{SSIM}(x, y)
= \frac{(2\mu_x \mu_y + C_1)(2\sigma_{xy} + C_2)}
       {(\mu_x^2 + \mu_y^2 + C_1)(\sigma_x^2 + \sigma_y^2 + C_2)}.
\end{equation}

The global structural similarity index is typically calculated as the mean of the \ac{SSIM} of sub-windows composing the image:
\begin{equation}
  \text{MSSIM}(X,Y) = \frac {1}{M} \sum_j \text{SSIM}(x_j,y_j)
  \label{eq:MSSIM}
 \end{equation}
where $X$ and $Y$ are the reference and the distorted images, respectively, $x_j$ and $y_j$ are the image contents at the $j$-th local window, and $M$ is the number of local windows in the image~\cite{wang2004image}.
While \ac{SSIM} is more perceptually meaningful than \ac{PSNR}, it also requires local-window statistics from both the reference and distorted signals and the calculation of the covariance based on pixel values of both original and distorted image.
%MM last part added
%LT good addition.

With the advent of Content Adaptive Encoding~\cite{li2016vmaf}, perceptual quality metrics have evolved from simple monitoring tools to deciding factors in modern video encoding pipelines. Perceptual metric evaluation is now embedded in rate-distortion optimization, bitrate ladder construction, and convex hull selection.
In these contexts, the same content is encoded repeatedly across many bitrates, resolutions, codecs, and encoder configurations. Although standard \ac{SSIM} is lightweight compared with modern encoding workloads, repeated full-reference evaluation still introduces compute and data-dependency costs at scale. Reducing these dependencies is therefore useful when source-derived statistics can be computed once, cached, and reused across multiple encodes of the same content.

This paper establishes a still-image 
%MM we have video in the title however
%LT yes, we should probably change the title
formulation for broader video-pipeline applications, using variance- and standard-deviation-based models to approximate \ac{SSIM} from global \ac{MSE} and reference-derived local statistics, without requiring local statistics from the distorted image.

Both approaches are validated on the Kodak and Xiph Subset1 datasets across a range of JPEG quality levels, demonstrating substantial accuracy gains over the global \ac{MSE} baseline and extending the framework of~\cite{martini2025ssim_psnr} to the more constrained single-image scenario.

\section{Related work}
\label{sec:related}

Zinner et al.~\cite{zinner2010towards} were among the early contributors to integrate perceptual quality metrics into multimedia systems. They relied on empirical mappings from \ac{PSNR} and \ac{SSIM} values to subjective quality categories. No analytical relationship was derived. Focusing on quality reduction caused by packet losses, in \cite{reibman2007characterizing} the authors used a quite simple relationship for block-based SSIM as a function of MSE, information on original and impaired image, and constant terms. The works in \cite{hore2010image,hore2013there} further analyzed the relationship and analytical expressions and approximations were provided for different use cases, requiring some joint processing of original and compressed image, beyond what required for PSNR. In \cite{dosselmann2011comprehensive} the authors, while discussing the validity of SSIM as quality metric, also showed that the index is directly related to the mean squared error, compared the two metrics statistically and derived a pair of functions that algebraically connect the two via means and mean square values of original and impaired images. Information on both original and impaired image is requested by their formulation in addition to MSE.

In \cite{tan2013perceptually} the authors developed a perceptually relevant MSE-based image quality metric. In doing so, they assume an additive error model and independence between signal and error. This metric is also adopted in \cite{yeo2013rate} and further tested in~\cite{martini2025ssim_psnr}, where it was shown that this estimation is less accurate than the methods proposed for \ac{DCT}-compressed images. Similar limitations were reported in~\cite{li2021joint}. We build here on top of the recent work \cite{martini2025ssim_psnr}, where a simple relationship is established between SSIM and local PSNR / MSE for DCT-based compressed images and videos.

\section{Proposed approach}
\label{sec:proposed}

The standard SSIM computation uses statistics computed over local overlapping windows of the original image ($X$) and the impaired image ($Y$). We consider the case where $Y$ is unavailable for local-window analysis: the only information about the distortion is the global frame-level MSE ($\text{MSE}_G$) between $X$ and $Y$.

Avoiding local-window analysis of \(Y\) has concrete advantages in modern video pipelines. Source-derived constants can be computed once and amortized across every encode of the same source. Because these constants are independent of the encode, they can also be computed ahead of time or in parallel with encoding.
The proposed approximation therefore reduces repeated local full-reference analysis to the combination of a reusable source-derived scalar and a global distortion measure already available from the encode.

Martini~\cite{martini2023simple,martini2025ssim_psnr} previously addressed a related problem, showing that local \ac{SSIM} values can be approximated by the local MSE ($\text{MSE}_{\ell}$) and the local variance of the source ($\sigma^2_x$)
\begin{equation}
\label{eq:ssim_local_mse}
\text{SSIM}(x_j, y_j) \approx 1 - \frac{\text{MSE}_{\ell}}{2\sigma^2_x + C_2}.
\end{equation}
Here $C_2$ is a standard \ac{SSIM} stabilization constant. The approximation reduces per-window \ac{SSIM} computation to a single ratio.

In this work, we consider a more restrictive case, where the $\text{MSE}_{\ell}$ is unavailable and only $\text{MSE}_{G}$ is known. Since SSIM operates on local windows, $\text{MSE}_{G}$ is insufficient to approximate $\text{MSE}_{\ell}$, due to the lack of spatial context.

To resolve this, we model \(MSE_\ell\) from local source activity.
In transform-based coding, quantization error is not spatially uniform after inverse transformation: smooth regions often retain lower absolute error, while textured or edge-rich regions can absorb larger errors. Local variance is therefore used as a first-order proxy for redistributing global distortion across SSIM windows.

An approximation of $\text{MSE}_{\ell}$ must satisfy
\begin{equation}
\label{eq:scale}
\text{MSE}_{G} = \mathbb{E}[\text{MSE}_{\ell}].
\end{equation}
This leads to a variance-normalized model such that
\begin{equation}
\label{eq:normalized}
\text{MSE}_{\ell} \approx \text{MSE}_{G} \cdot \frac{\sigma^2_x}{\mathbb{E}[\sigma^2_x]}.
\end{equation}
This distributes global distortion according to local activity.

Empirical analysis of the previous equation revealed over-allocation of error to high variance regions. We soften the assumption to a sublinear relationship and introduce a shaping exponent $\beta \in [0,1]$
\begin{equation}
\label{eq:sublinear}
\text{MSE}_{\ell} \approx \text{MSE}_{G} \cdot \frac{(\sigma^2_x + \epsilon)^{\beta}}{\mathbb{E}[(\sigma^2_x + \epsilon)^\beta]}.
\end{equation}
In the previous equation, $\epsilon=1e^{-6}$.

Optimizing $\beta$ over the Kodak and Xiph Subset1 datasets resulted in values of 0.46 and 0.48, respectively. While broader validation would be needed to establish generality, this consistency suggests a sublinear relationship close to a square root, motivating a parameter-free model based on standard deviation $(\sigma_x)$ rather than variance. This gives:
\begin{equation}
\label{eq:stdbased}
\text{MSE}_{\ell} \approx \text{MSE}_{G} \cdot \frac{\sigma_x + \epsilon}{\mathbb{E}[\sigma_x + \epsilon]}.
\end{equation}

Compared with variance, standard deviation compresses the dynamic range of local activity and reduces the dominance of highly textured regions, leading to a more balanced first-order model of artifact distribution.

The local SSIM is hence obtained based on global image MSE (or PSNR) and  variance of one image (e.g., the original uncompressed one).
\begin{equation}
\label{eq:ssim_local_mse_final}
\text{SSIM}(x_j, y_j) \approx 1 - \frac{\text{MSE}_{G} \cdot (\sigma_x + \epsilon)}{\mathbb{E}[\sigma_x + \epsilon]\,(2\sigma^2_x + C_2)}.
\end{equation}

The global SSIM is computed by averaging over all the windows of the frame:
\begin{equation}
  \text{MSSIM}(X,Y) \approx \frac{1}{M} \sum_j \left(1 - \frac  {\text{MSE}_{G}  \cdot (\sigma_x + \epsilon)} {\mathbb{E}[\sigma_x + \epsilon]\,(2\sigma_x^2 + C_2)} \right).
  \label{eq:MSSIMfinal}
 \end{equation}
Since $\text{MSE}_{G}$ is a constant with respect to $j$, it factors out
 \begin{equation}
  \text{MSSIM}(X,Y) \approx 1 - \text{MSE}_{G} \cdot \underbrace{\frac{1}{M} \sum_j \left(\frac{\sigma_x + \epsilon}{\mathbb{E}[\sigma_x + \epsilon]\,(2\sigma_x^2 + C_2)} \right)}_{k}.
  \label{eq:MSSIMSimplified}
 \end{equation}
Here $k$ is a source-only scalar. The mean \ac{SSIM} of a frame reduces to a product of a source-dependent constant $k$ and a distortion scalar~$\text{MSE}_{G}$. No per-window SSIM computation is required.

\section{Numerical results}
\label{sec:results}

\begin{figure*}[t]
\centering
\begin{subfigure}{.32\textwidth}
    \centering
    \begin{tikzpicture}
\begin{axis}[
    width=\columnwidth,
    height=\columnwidth,
    xlabel={JPEG quality},
    ylabel={SSIM (Y)},
    ymin=0.7, ymax=1.0,
    legend to name=legendref,
    legend columns=5,
    grid=both,
]

\addplot[blue, mark=*, mark size=1.6pt, line width=0.7pt,]
    table[x=quality, y=Local MSE, col sep=comma]{kodak_ssim_vs_quality.csv};
\addlegendentry{Local MSE}

\addplot[orange, mark=*, mark size=1.6pt, line width=0.7pt,]
    table[x=quality, y=Global MSE, col sep=comma]
    {kodak_ssim_vs_quality.csv};
\addlegendentry{Global MSE}

\addplot[green!60!black, mark=*, mark size=1.6pt, line width=0.7pt,]
    table[x=quality, y=Sublinear variance, col sep=comma]
    {kodak_ssim_vs_quality.csv};
\addlegendentry{Sublinear variance}

\addplot[red, mark=*, mark size=1.6pt, line width=0.7pt,]
    table[x=quality, y=Std-based, col sep=comma]
    {kodak_ssim_vs_quality.csv};
\addlegendentry{Std-based}

\addplot[black, thick, mark=none] table[x=quality, y=SSIM, col sep=comma]
{kodak_ssim_vs_quality.csv};
\addlegendentry{SSIM}

\end{axis}
\end{tikzpicture}
    \vspace{-1.5em}
    \caption{Kodak}
    \label{fig:kodak_ssim_vs_quality}
\end{subfigure}
\hfill
\begin{subfigure}{.32\textwidth}
    \centering
    \begin{tikzpicture}
\begin{axis}[
    width=\columnwidth,
    height=\columnwidth,
    xlabel={JPEG quality},
    ylabel={SSIM (Y)},
    ymin=0.66, ymax=1.0,
    legend style={draw=none, fill=none, opacity=0},
    grid=both,
]

\addplot[blue, mark=*, mark size=1.6pt, line width=0.7pt,]
    table[x=quality, y=Local MSE, col sep=comma]{subset1_ssim_vs_quality.csv};
\addlegendentry{Local MSE}

\addplot[orange, mark=*, mark size=1.6pt, line width=0.7pt,]
    table[x=quality, y=Global MSE, col sep=comma]
    {subset1_ssim_vs_quality.csv};
\addlegendentry{Global MSE}

\addplot[green!60!black, mark=*, mark size=1.6pt, line width=0.7pt,]
    table[x=quality, y=Sublinear variance, col sep=comma]
    {subset1_ssim_vs_quality.csv};
\addlegendentry{Sublinear variance}

\addplot[red, mark=*, mark size=1.6pt, line width=0.7pt,]
    table[x=quality, y=Std-based, col sep=comma]
    {subset1_ssim_vs_quality.csv};
\addlegendentry{Std-based}

\addplot[black, thick, mark=none] table[x=quality, y=SSIM, col sep=comma]
{subset1_ssim_vs_quality.csv};
\addlegendentry{SSIM}

\end{axis}
\end{tikzpicture}
    \vspace{-1.5em}
    \caption{Subset1}
    \label{fig:subset1_ssim_vs_quality}
\end{subfigure}
\hfill
\begin{subfigure}{.32\textwidth}
    \centering
    \begin{tikzpicture}
\begin{axis}[
    width=\columnwidth,
    height=\columnwidth,
    xlabel={JPEG quality},
    ylabel={SSIM (Y) Delta (\%) },
    ymin=-2, ymax=2,
    legend style={draw=none, fill=none, opacity=0},
    grid=both,
]

\addplot[blue, mark=*, mark size=1.6pt, line width=0.7pt,]
    table[x=quality, y=Local MSE (bound), col sep=comma]{subset1_ssim_delta.csv};
\addlegendentry{Local MSE (bound)}

\addplot[green!60!black, mark=*, mark size=1.6pt, line width=0.7pt,]
    table[x=quality, y=Sublinear variance, col sep=comma]
    {subset1_ssim_delta.csv};
\addlegendentry{Sublinear variance}

\addplot[red, mark=*, mark size=1.6pt, line width=0.7pt,]
    table[x=quality, y=Std-based, col sep=comma]
    {subset1_ssim_delta.csv};
\addlegendentry{Std-based}

\end{axis}
\end{tikzpicture}
    \vspace{-1.5em}
    \caption{Approximation error (Subset1)}
\label{fig:subset1_ssim_delta}
\end{subfigure}
\vspace{4pt}
\ref{legendref}
\caption{Average SSIM versus JPEG quality averaged across images in the Kodak (a) and Subset1 (b) datasets, and difference between SSIM and SSIM approximations as a function of JPEG quality on Subset1 (c).}
\label{fig:ssim_vs_quality}
\vspace{-10pt}
\end{figure*}
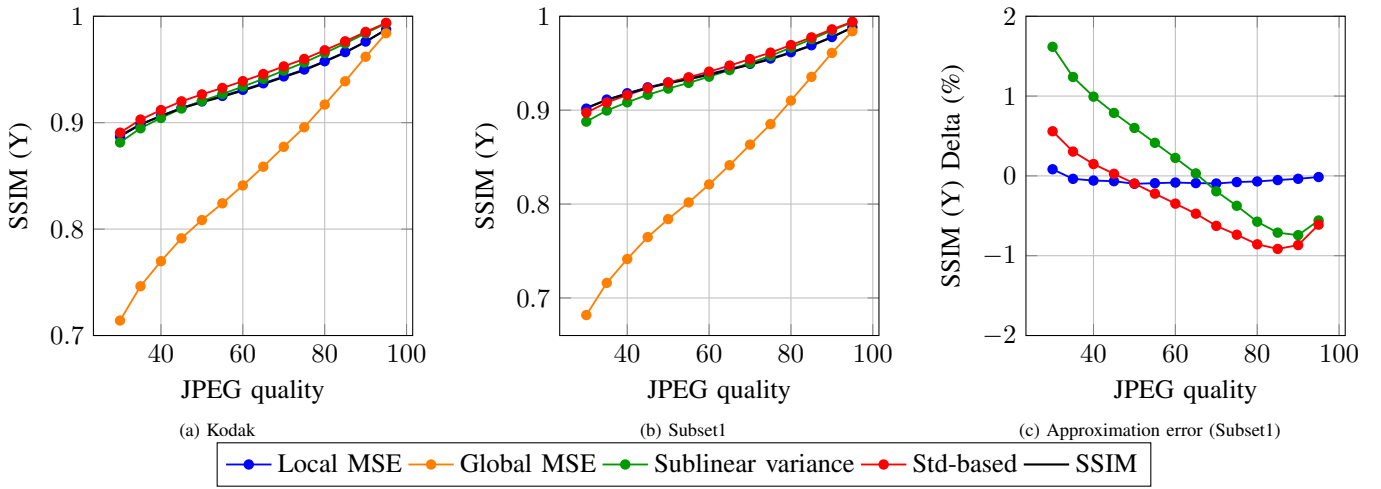

Experiments use the Kodak Lossless True Color Image Suite~\cite{kodak_dataset} and the Xiph.org Foundation Subset1 still-image test set~\cite{subset1_dataset}, comprising 24 and 50 high-quality uncompressed images, respectively.

\ac{SSIM} values were computed with scikit-image~\cite{scikit-image} using the default $7\times7$ windows, matching the standard formulation~\cite{wang2004image}; the proposed approximations are not implementation-specific. The code is available as open-source software~\cite{trudeau2026approxssimate}.

JPEG is used as a representative example of \ac{DCT}-based compression. Compression was performed with Pillow~\cite{pillow}, using quality values from 35 to 95 in steps of 5 to cover practical lossy operating points while avoiding both near-lossless saturation and extremely low-quality regimes where artifacts become less representative of typical usage.

We compare four approximations: Martini's local-\ac{MSE} method~\cite{martini2025ssim_psnr}, a global-\ac{MSE} baseline, and the proposed sublinear-variance and standard-deviation models. The proposed models use source variance over \(7\times7\) windows to redistribute global \ac{MSE} locally.

\Cref{fig:kodak_ssim_vs_quality,fig:subset1_ssim_vs_quality} illustrate the behavior of the proposed SSIM approximations as a function of JPEG quality, compared with the reference SSIM values (black curve). All SSIM values are averaged over the images of the corresponding dataset.

Across the full range of evaluated qualities, the proposed approaches consistently provide more accurate SSIM estimates than the global MSE-based approximation. Notably, the approximations preserve the relative ordering of quality levels, indicating that they remain suitable for comparative evaluation.
These results demonstrate that weighting the global MSE using local signal statistics yields SSIM estimates close to those obtained using local MSE, while requiring local statistics only from the reference image, together with the global distortion measure already available from the encode.

Moreover, \Cref{fig:subset1_ssim_delta} complements \cref{fig:kodak_ssim_vs_quality,fig:subset1_ssim_vs_quality} by providing a closer view of the approximation error for the local MSE, sublinear variance, and the standard-deviation-based approach. The y-axis shows the difference between reference SSIM and the corresponding approximation, while the x-axis represents the JPEG quality level. All results are averaged over the images of Subset1. The curve corresponding to local MSE highlights the residual error introduced by the approximation proposed in \cite{martini2025ssim_psnr}.

At lower JPEG quality levels, the sublinear variance-based approximation shows larger deviations, consistent with an over-allocation of distortion to highly textured regions when artifacts are severe. The standard-deviation-based model reduces this effect by compressing the dynamic range of local activity, leading to a more balanced redistribution of global error. These results also indicate a limitation of the proposed model: at very low quality, distortion becomes increasingly artifact-dependent and less well described by source-only local statistics. Nevertheless, over the evaluated practical quality range, the approximation error remains small relative to the overall SSIM scale, and both proposed methods preserve the monotonic relationship between JPEG quality and SSIM.

\section{Conclusion}
\label{sec:conclusion}

This paper showed that, for \ac{DCT}-based compressed images, \ac{SSIM} can be approximated from global \ac{PSNR} (or \ac{MSE}) and reference-derived local statistics. On JPEG-compressed Kodak and Xiph Subset1 images, the approximation error remains below 1\% across typical quality levels, and the optimized exponent $\beta$ is consistently close to 0.5, motivating a parameter-free standard-deviation-based formulation. The present validation is limited to JPEG still images; 
%MM: We can add to justify the title something like: "While based on the preliminary results from Martini[]  we expect that the presented model performs similary well for video, we leave to future work the evaluation for...
%LT Hopefully, we can change the title of the paper. Also feel free to change the conclusion as you see fit.
future work should evaluate modern image and video codecs, compare runtime against local-\ac{MSE} approximations, and quantify the benefit of amortizing source-derived statistics across repeated video encodes.

\bibliographystyle{IEEEbib}
\bibliography{refs}

@misc{trudeau2026approxssimate,
  author       = {Trudeau, Luc and Martini, Maria G.},
  title        = {{ApproxSSIMate}: Estimating {SSIM} from {PSNR}/{MSE} and Reference-Derived Statistics},
  year         = {2026},
  howpublished = {\url{https://github.com/luctrudeau/approxssimate}},
  note         = {Open-source software, version 0.1.0, BSD-2-Clause license}
}

@inproceedings{li2016vmaf,
  author    = {Li, Zhi and Aaron, Anne and Katsavounidis, Ioannis and 
               Moorthy, Anush and Manohara, Megha},
  title     = {Toward A Practical Perceptual Video Quality Metric},
  booktitle = {Netflix Technology Blog},
  year      = {2016},
  url       = {https://netflixtechblog.com/toward-a-practical-perceptual-video-quality-metric-653f208b9652}
}

@misc{kodak_dataset,
  title        = {Kodak Lossless True Color Image Suite},
  author       = {{Eastman Kodak Company}},
  howpublished = {\url{https://r0k.us/graphics/kodak/}},
  year         = {1999},
  note         = {Accessed: 2026-01-}
}

@article{martini2025ssim_psnr,
  author  = {Martini, Maria G.},
  title   = {Measuring Objective Image and Video Quality: On the Relationship Between {SSIM} and {PSNR} for {DCT}-Based Compressed Images},
  journal = {IEEE Transactions on Instrumentation and Measurement},
  year    = {2025},
  volume  = {74},
  pages   = {1--13},
  doi     = {10.1109/TIM.2025.3529045}
}

@misc{pillow,
  title        = {Pillow: Python Imaging Library},
  author       = {Clark, Alex and contributors},
  howpublished = {\url{https://python-pillow.org/}},
  note         = {Accessed: 2026-01}
}

@article{scikit-image,
  title   = {scikit-image: image processing in Python},
  author  = {van der Walt, Stefan and Sch{\"o}nberger, Johannes L. and
             Nunez-Iglesias, Juan and Boulogne, Fran{\c{c}}ois and
             Warner, Joshua D. and Yager, Neil and Gouillart, Emmanuelle and
             Yu, Tony and the scikit-image contributors},
  journal = {PeerJ},
  volume  = {2},
  pages   = {e453},
  year    = {2014},
  doi     = {10.7717/peerj.453}
}

@misc{subset1_dataset,
  title        = {Subset1 Image Dataset},
  author       = {{Xiph.org Foundation}},
  year         = {2012},
  howpublished = {\url{https://media.xiph.org/video/derf/subset1-y4m.tar.gz}},
  note         = {Accessed: 2026-01}
}

@article{hore2013there,
  title={Is there a relationship between peak-signal-to-noise ratio and structural similarity index measure?},
  author={Hor{\'e}, Alain and Ziou, Djemel},
  journal={IET Image Processing},
  volume={7},
  number={1},
  pages={12--24},
  year={2013},
  publisher={IET}
}

@inproceedings{hore2010image,
  title={Image quality metrics: {PSNR vs. SSIM}},
  author={Hore, Alain and Ziou, Djemel},
  booktitle={2010 20th International Conference on Pattern Recognition},
  pages={2366--2369},
  year={2010},
  organization={IEEE}
}

@article{wang2004image,
  title={Image quality assessment: from error visibility to structural similarity},
  author={Wang, Zhou and Bovik, Alan C and Sheikh, Hamid R and Simoncelli, Eero P},
  journal={IEEE Transactions on Image Processing},
  volume={13},
  number={4},
  pages={600--612},
  year={2004},
  publisher={IEEE}
}

@inproceedings{zinner2010towards,
  title={Towards {QoE} management for scalable video streaming},
  author={Zinner, Thomas and Abboud, Osama and Hohlfeld, Oliver and Hossfeld, Tobias and Tran-Gia, Phuoc},
  booktitle={21th ITC specialist seminar on multimedia applications-traffic, performance and QoE},
  pages={64--69},
  year={2010},
  organization={Citeseer}
}

@article{yeo2013rate,
  title={On rate distortion optimization using {SSIM}},
  author={Yeo, Chuohao and Tan, Hui Li and Tan, Yih Han},
  journal={IEEE Transactions on Circuits and Systems for Video Technology},
  volume={23},
  number={7},
  pages={1170--1181},
  year={2013},
  publisher={IEEE}
}

@article{tan2013perceptually,
  title={A perceptually relevant {MSE}-based image quality metric},
  author={Tan, Hui Li and Li, Zhengguo and Tan, Yih Han and Rahardja, Susanto and Yeo, Chuohuo},
  journal={IEEE Transactions on Image Processing},
  volume={22},
  number={11},
  pages={4447--4459},
  year={2013},
  publisher={IEEE}
}

@article{dosselmann2011comprehensive,
  title={A comprehensive assessment of the structural similarity index},
  author={Dosselmann, Richard and Yang, Xue Dong},
  journal={Signal, Image and Video Processing},
  volume={5},
  pages={81--91},
  year={2011},
  publisher={Springer}
}

@inproceedings{reibman2007characterizing,
  title={Characterizing packet-loss impairments in compressed video},
  author={Reibman, Amy R and Poole, David},
  booktitle={2007 IEEE International Conference on Image Processing},
  volume={5},
  pages={V--77},
  year={2007},
  organization={IEEE}
}

@inproceedings{martini2023simple,
  title={{A Simple Relationship Between SSIM and PSNR for DCT-Based Compressed Images and Video: SSIM as Content-Aware PSNR}},
  author={Martini, Maria},
  booktitle={2023 IEEE 25th International Workshop on Multimedia Signal Processing (MMSP)},
  pages={1--5},
  year={2023},
  organization={IEEE}
}

@article{li2021joint,
  title={Joint optimization for {SSIM-based CTU}-level bit allocation and rate distortion optimization},
  author={Li, Yang and Mou, Xuanqin},
  journal={IEEE Transactions on Broadcasting},
  volume={67},
  number={2},
  pages={500--511},
  year={2021},
  publisher={IEEE}
}

\end{document}